# Faster Approximate Distance Queries and Compact Routing in Sparse Graphs *

Rachit Agarwal[†] P. Brighten Godfrey[‡] Sariel Har-Peled[§]

January 11, 2012

**Abstract**

A distance oracle is a compact representation of the shortest distance matrix of a graph. It can be queried to approximate shortest paths between any pair of vertices. Any distance oracle that returns paths of worst-case stretch $(2k-1)$ must require space $\Omega(n^{1+1/k})$ for graphs of $n$ nodes. The hard cases that enforce this lower bound are, however, rather dense graphs with average degree $\Omega(n^{1/k})$.

We present distance oracles that, for sparse graphs, substantially break the lower bound barrier at the expense of higher query time. For any $1 \leq \alpha \leq n$, our distance oracles can return stretch 2 paths using $O(m + n^2/\alpha)$ space and stretch 3 paths using $O(m + n^2/\alpha^2)$ space, at the expense of $O(\alpha m/n)$ query time. By setting appropriate values of $\alpha$, we get the first distance oracles that have size linear in the size of the graph, and return constant stretch paths in non-trivial query time. The query time can be further reduced to $O(\alpha)$, by using an additional $O(m\alpha)$ space for all our distance oracles, or at the cost of a small constant additive stretch.

We use our stretch 2 distance oracle to present the first compact routing scheme with worst-case stretch 2. Any compact routing scheme with stretch less than 2 must require linear memory at some nodes even for extremely sparse graphs; our scheme, hence, achieves the optimal stretch with non-trivial memory requirements. Moreover, supported by large-scale simulations on graphs including the AS-level Internet graph, we argue that our stretch-2 scheme would be simple and efficient to implement as a distributed compact routing protocol.

†University of Illinois, Urbana, IL 61801. Email: agarwa16@illinois.edu.

‡University of Illinois, Urbana, IL 61801. Email: pbg@illinois.edu.

§University of Illinois, Urbana, IL 61801. Email: sariel@illinois.edu.

# 1 Introduction

A distance oracle is a compact representation of the shortest distance matrix of a graph. It can be queried to retrieve distances and paths (of corresponding length) between any pair of nodes in the graph. Besides their fundamental connection to the all-pair shortest path problem, there are two main applications of distance oracles. First, techniques from distance oracles have been applied in designing compact routing schemes [14], where routers must require limited memory to store forwarding tables and yet route along short paths. Second, distance oracles have been used to analyze large scale social networks [10] – efficiently computing shortest distances and paths between users in social networks is both important and non-trivial. These networks often contain millions of nodes and billions of edges, making it expensive (if not impossible) to store the shortest distance matrix on machines with limited random access memory. To make computations feasible, distance oracles with smaller size and bounded small error on distances returned are desired.

Indeed, a fundamental trade-off in constructing a distance oracle is between its size and its *stretch*: the worst-case ratio of the distance returned by the distance oracle to the actual shortest distance between the two vertices. For general graphs, the optimal[1] space/stretch trade-off was achieved by Thorup and Zwick [15]: their distance oracle, for any graph with $n$ vertices and for any integer $k \geq 2$, is of size $O(kn^{1+1/k})$ and returns paths with stretch $2k-1$ in time $O(k)$. However, the hard instances for the matching lower bound are rather dense graphs, with average degree $\Omega(n^{1/k})$. For instance, to prove a space lower bound of $\Omega(n^2)$ for stretch 2, their proof uses a graph with $\Theta(n^2)$ edges; for stretch 3, the proof uses a graph with $\Omega(n^{3/2})$ edges. The lower bound essentially states that there exist graphs that are incompressible: if a certain stretch is desired, then the size of the distance oracle is lower bounded by the number of edges in the specially-constructed graph.

Thus, classic distance oracle results may be quite far from optimal for sparse graphs: graphs with low average degree $\Delta$. The notion of sparsity is a little tricky – for stretch 2, graphs with average degree $\Delta = o(n)$ are said to be sparse; for any integer $k \geq 2$ and stretch at most $2k-1$, graphs with $\Delta = o(n^{1/k})$ are said to be sparse. This is of key interest since real-world graphs are sparse, with degrees much closer to logarithmic than polynomial in $n$. For instance, letting $\Delta = c\log_2 n$, empirically, $c \approx 0.6$ for an AS-level map of the Internet [13], $c \approx 0.4$ for a router-level map of the Internet [13], and $c \approx 1.34, 0.65, 1.21, 5.10, 29.9$ for social networks Cyworld, Testimonial, Orkut, MySpace, and Facebook, respectively [2,7].

## 1.1 Our contributions

**Distance oracles.** This paper presents distance oracles that, for sparse graphs, substantially break the classic space/stretch trade-off barrier, albeit at the cost of increased query time. For instance, in dense graphs, retrieving distances of stretch 2 and 3 in constant time requires space $\Theta(n^2)$ and $\Theta(n^{3/2})$ respectively [15]; larger query time can not help reduce space and/or stretch. We present distance oracles that have size linear in the size of the graph and return stretch 2 and 3 distances in sub-linear query time; this significantly improves upon the earlier known constructions for the case of real-world networks where average degree is logarithmic in the number of nodes. Moreover, we demonstrate that our approach allows a surprisingly large fraction of source-destination pairs to retrieve exact distances and *shortest* paths.

More specifically, we introduce several new distance oracles which respectively improve stretch and space in comparison with the distance oracle of Thorup and Zwick [15]. Let $1 \leq \alpha \leq n$ and $k$ be any positive integer. Then, for weighted undirected graphs with $n$ vertices and average degree $\Delta$, we present distance oracles that return stretch 2 distances using space $O(n\Delta + n^2/\alpha)$ and stretch $(4k-1)$ distances using space $O(n\Delta + (n/\alpha)^{(1+1/k)})$; both these distance oracles require $O(\alpha\Delta)$ query time. The query time can be further reduced to $O(\alpha)$ using an additional $O(n\alpha\Delta)$ space or a small constant additive factor (see Table 1 in §2).

[1]For $k = 4$ and $k \geq 6$, the lower bound relies on a conjecture of Erdős.

For example, for the realistic case of $\Delta = \Theta(\text{polylog}(n))$, special cases of our two results yield schemes for retrieving stretch 2 distances using space $\tilde{O}(n^{3/2})$, and stretch 3 distances using space $\tilde{O}(n)$, at the expense of $\tilde{O}(\sqrt{n})$ query time. Out of theoretical interest, we note that our distance oracles highlight the fact that in the regime of sparse graphs, for any fixed stretch, there may be an infinite number of *pareto optimal* design points – one can smoothly trade-off the query time to reduce the space requirements; in contrast, for dense graphs, there is exactly one optimal design point for any fixed stretch.

**Compact routing.** Thorup and Zwick [14] designed compact routing schemes for their distance oracles. Their scheme requires $\tilde{O}(\sqrt{n})$ memory at each node in the network and routes along paths that have stretch 3. No compact routing schemes are known for stretch less than 3 for general graphs; in fact, it is known that even for extremely sparse graphs, any compact routing scheme that route along paths of stretch less than 2 must use $\Omega(n)$ memory at some nodes in the network [5]. Hence, all we can hope for is compact routing schemes with stretch 2 and larger.

For graphs with average degree $\Delta = o(n)$, we present the first compact routing scheme with the optimal stretch. The scheme requires $O(\alpha\Delta + n/\alpha)$ memory at each router and route along paths of worst-case stretch 2. Besides being the first compact routing scheme (for general graphs) with provable optimal stretch, our compact routing scheme has a particular property: it can be implemented on top of any implementation of Thorup-Zwick scheme using a *handshaking* scheme – a surprisingly lightweight end-to-end exchange of a very few packets – and a small amount of processing to set up a new end-to-end connection with worst-case stretch 2. Using a distributed protocol [11] to construct Thorup-Zwick compact routing scheme (with appropriately setting the parameters), we get a distributed name-independent compact routing scheme for our distance oracles with roughly the same space requirements.

In summary, our results represent a step towards characterizing the space/stretch/time trade-off for approximate distance queries in sparse graphs, and yield a simple, practical way to improve stretch in compact routing protocols. We complement our theoretical results with extensive simulations on empirical networks. Interestingly, we find that in the Internet AS-level topology, our stretch-2 scheme finds *shortest* paths for 99.98% of the source-destination pairs – compared with 34.4% using [15].

**Roadmap.** We start our discussion with related work (§2) and setting up the notation used in the paper (§3). In §4, we give an overview of the main techniques used in design of our distance oracles. One of the challenges in designing distance oracles with the minimalistic assumption on number of edges in the graph is to handle skewed degree distribution; in §5, we show how to handle this challenge. In particular, we prove that in the context of designing distance oracles, average-degree-bounded graphs are no harder than maximum-degree-bounded graphs; this allows us to restrict our attention to maximum-degree-bounded graphs in the rest of the paper. We present distance oracles for stretch 2 in §6 and for stretch 3 and higher in §7. We show how to improve the space/query time trade-off in our distance oracles using a small (constant) additive stretch in §8. In §9, we design compact routing schemes for our distance oracles and describe how to implement our schemes in a distributed fashion. We evaluate the performance of our distance oracles and compact routing schemes on a number of synthetic and real-world graph datasets; the evaluation results and analysis is presented in §10. Finally, we close the paper by outlining a number of open problems in §11.

**Bibliographic note.** This paper extends and improves upon the results presented in the conference version of this paper [1]. In particular, to bound the worst-case stretch, this version presents conditions (Lemma 1) that are tighter than those used in the conference version, and use the new conditions to improve the query time without increasing the size of the distance oracles (updated §6 and §7). The improvement in query time also leads to improved compact routing schemes (updated §9). In addition, this version also presents distance oracles for additive stretch (new §8). Finally, we have significantly simplified the presentation and proofs when compared to the conference version (essentially re-written §6 and §7).

## 2 Related Work

In this section, we discuss the known lower and upper bounds for the space/stretch trade-off in the approximate distance query problem for the regime of sparse graphs.

**Lower bounds.** For general graphs, Thorup and Zwick [15] showed (subject to a conjecture of Erdős) that achieving (integer) stretch $(2k-1)$ requires $\Omega(kn^{1+1/k})$ space. Their proof is information-theoretic, essentially showing that for any constant stretch, there exist graphs that require storing as many bits as the number of edges in the graph. For example, proving that stretch 2 requires $\Omega(n^2)$ space uses a graph with $\Theta(n^2)$ edges; for stretch 3, the proof uses a graph with $\Theta(n^{3/2})$ edges.

There is no hope of this proof technique being helpful in the sparse case. In particular, for graphs with $m = n\Delta$ edges, this technique will only show that achieving any constant stretch value requires $\Omega(n\Delta)$ bits. This much space is entirely acceptable for sparse graphs, and in fact, *can* permit retrieval of shortest paths, simply by storing the original graph and running Dijkstra's algorithm for each query. Of course, this takes time $O(n\Delta)$ per query. Thus, in the context of distance oracles with super-constant query time, the cases of dense and sparse graphs are quite different. In the dense case the key is to *compress* the graph while ensuring that sufficient information remains to return low-stretch distances. In the sparse case the graph need not be compressed, but the trade-off with *query time* becomes critical.

Very little is known about this trade-off space for sparse graphs. First, Sommer et al. [12] show that any distance oracle that returns stretch $t$ paths in time $\alpha$ requires space $n^{1+\Omega(1/\alpha t)}$. For distance oracles with constant query time, this gives a lower bound of space $n^{1+\Omega(1/t)}$ for any stretch $t$. However, if we allow $\Omega(\log n)$ query time, their result implies a trivial lower bound of $\Omega(n \log n)$ for any constant stretch. Second, Pătraşcu and Roditty [9] prove that if a widely believed conjecture about the hardness of set intersection queries holds, then retrieving stretch 2 paths in constant time requires a distance oracle of size $\Omega(n\sqrt{n\Delta})$. For the case of $\Omega(\log n)$ query time, as in our schemes, no non-trivial lower bounds are known.

For stretch 3 and larger, the lower bounds for distance oracles also hold for compact routing schemes [14]; consequently, these are tight only for dense graphs. It is shown in [4–6] that any compact routing scheme with stretch less than 2 must require $\Omega(n \log n)$ memory at some nodes in the network – this bound holds even for extremely sparse graphs. The compact routing scheme for our distance oracle, hence, achieves the optimal stretch with non-trivial memory requirements at each node.

**Upper bounds.** A detailed comparison of our results with previously known upper bounds on distance oracles for general graphs is presented in Table 1. Very recently, Pătraşcu and Roditty [9] obtained a distance oracle that returns stretch 2 paths in constant time with $O(\Delta^{1/3}n^{5/3})$ space. These queries are faster than our stretch-2 scheme, but the distance oracle has larger size for $\alpha > (n/\Delta)^{1/3}$. For general sparse graphs, no other results are known.

For unweighted graphs, the only known $o(n^2)$ size distance oracle with approximation ratio 2 is again due to the recent result of Pătraşcu and Roditty [9]. Their distance oracle requires space $O(n^{5/3})$ and returns approximate distance in constant time. As earlier, our distance oracle requires lesser space but higher query time when compared to their distance oracle.

In terms of upper bounds for compact routing schemes, we note that the only known results are by Thorup and Zwick [14] for stretch 3. No compact routing schemes with worst-case stretch less than 3 are known. Although we believe that it may be possible to design compact routing schemes for the distance oracle of Pătraşcu and Roditty [9], it is not clear whether this can be done in a distributed fashion. Our compact routing schemes, on the other hand, can be constructed in a distributed fashion and have worst-case stretch bounded by 2.

**Table 1:** Upper bounds for distance oracles for general undirected graphs. The $(\alpha, \beta)$ in column 3 denotes multiplicative $\alpha$ and additive $\beta$ approximation ratio. $\Delta$ denotes the *average* degree of the graph. Distance oracles with additive stretch are for unweighted graphs.

| Reference | Space | Stretch | Query Time | Remarks |
|---|---|---|---|---|
| [15] | $O(n^2)$ | 2 | $O(1)$ | Optimal space if $m = \Theta(n^2)$ |
| [9] | $O(n^{5/3}\Delta^{1/3})$ | 2 | $O(1)$ | |
| **§6** | $O(n\Delta + n^2/\alpha)$ | 2 | $O(\alpha\Delta)$ | $1 \leq \alpha \leq n$ |
| **§6** | $O(n\alpha\Delta + n^2/\alpha)$ | 2 | $O(\alpha)$ | $1 \leq \alpha \leq n$ |
| [15] | $O(n^{3/2})$ | 3 | $O(1)$ | Optimal space if $m = \Omega(n^{3/2})$ |
| **§7** | $O(n\Delta + n^2/\alpha^2)$ | 3 | $O(\alpha\Delta)$ | $1 \leq \alpha \leq n$ |
| **§7** | $O(n\alpha\Delta + n^2/\alpha^2)$ | 3 | $O(\alpha)$ | $1 \leq \alpha \leq n$ |
| [15] | $O(kn^{1+1/k})$ | $2k-1$ | $O(k)$ | Optimal space if $m = \Omega(n^{1+1/k})$ |
| **§7** | $O(n\Delta + (n/\alpha)^{(1+1/k)})$ | $4k-1$ | $O(\alpha\Delta)$ | $1 \leq \alpha \leq n$ |
| **§7** | $O(n\alpha\Delta + (n/\alpha)^{(1+1/k)})$ | $4k-1$ | $O(\alpha)$ | $1 \leq \alpha \leq n$ |
| [15] | $O(n^{3/2})$ | $(3,0)$ | $O(1)$ | |
| [9] | $O(n^{5/3})$ | $(2,1)$ | $O(1)$ | |
| **§8** | $O(n\alpha + n^2/\alpha)$ | $(2,1)$ | $O(\alpha)$ | $1 \leq \alpha \leq n$ |
| **§8** | $O(n\alpha + n^2/\alpha^2)$ | $(3,2)$ | $O(\alpha)$ | $1 \leq \alpha \leq n$ |
| **§8** | $O(n\alpha + (n/\alpha)^{(1+1/k)})$ | $(4k-1, 2k)$ | $O(\alpha)$ | $1 \leq \alpha \leq n$ |

## 3 Notations and Definitions

Throughout the paper, we let $G = (V, E)$ be a connected, undirected graph with $n = |V|$ nodes and $m = |E|$ edges. Unless mentioned otherwise, $G$ is assumed to be weighted with each edge assigned a non-negative weight.

For any node $v \in V$, we denote by $N(v)$ the set of all the neighbors of $v$. For any set $V' \subset V$, we denote by $N(V')$ the set of all the neighbors of nodes in $V'$. We let $\deg(v)$ denote the number of neighbors of node $v$, that is, $\deg(v) = |N(v)|$. The graph is said to be maximum-degree-bounded (or, $\Delta$-degree bounded) if for all nodes $v \in V$, $\deg(v) \leq \Delta$. We say that the graph is average-degree-bounded graph (or, has average degree $\Delta$) if $2m/n \leq \Delta$.

For any pair of nodes $u, v \in V$, let $d(u,v)$ be the length of the shortest path between $u$ and $v$ in $G$ and let $\delta(u,v)$ be the length of the path returned by the distance oracle. The distance oracle is said to return stretch $t$ paths if for every pair of nodes $u, v \in V$, $d(u,v) \leq \delta(u,v) \leq t \cdot d(u,v)$.

We will let $L \subset V$ denote a distinguished set of "landmark" nodes chosen by our algorithms. For any node $v \in V$, we denote by $\ell(v)$ the nearest neighbor of $v$ in $L$ (i.e., the node $a \in L$ that minimizes $d(v,a)$, with ties broken arbitrarily). The **ball of** $v$, $B(v)$, is the set of nodes $w \in V$ for which $d(v,w) < d(v,\ell(v))$ and the **vicinity** of $v$ is $\Gamma(v) = B(v) \cup N(B(v))$.

# 4 Overview of our schemes

Our distance oracle for stretch 2 is conceptually similar to the stretch 3 distance oracle of Thorup and Zwick [15]. For a given graph, they construct a set of nodes, known as *landmarks*, such that each node has a landmark in its ball. The distance oracle stores, for each node, the distance to each node in its ball and to its closest landmark; the landmarks store distances to all nodes in the graph. When queried for distance between nodes $u$ and $v$, the query algorithm checks if $v$ is in ball of $u$. If it is, then the exact distance is returned using information stored in the distance oracle; if not, the distance $d(u, \ell(u)) + d(\ell(u), v)$ is returned, which by triangle inequality is at most of stretch 3.

Intuitively, the cases that attain worst-case stretch in their distance oracle are the ones for which the destination $v$ is just outside the ball of the source $u$. For such source-destination pairs, we exploit the idea of *ball-vicinity intersection*[2]. Upon receiving a query, we search for nodes in $B(u) \cap \Gamma(v)$. Finding such nodes takes some time; but if any such node $w$ exists, we can return the distance $d(u,w) + d(w,v)$ using information stored in the distance oracle. If $B(u) \cap \Gamma(v) = \emptyset$, the nodes must be relatively distant, giving us a lower bound on the exact distance between $u$ and $v$. Using this lower bound, we show that a path via the landmark node has stretch 2. We need to store the vicinities of the nodes for some of our distance oracles; but if the graph is sparse, we show that this does not increase the space requirements significantly.

The above distance oracle is of large space since it requires storing (a) shortest paths from the landmarks to all other nodes; and (b) the vicinities of every node. To avoid the first requirement, our distance oracles for stretch 3 and larger store the exact distances only between all pairs of landmarks. This uses significantly less space; for instance, in a graph with $n$ nodes, storing shortest paths between every pair of $\sqrt{n}$ landmarks requires space at most linear in the size of the graph. To overcome the second requirement, our distance oracles computes the vicinities *on the fly* during a query; we show that for sparse graphs, this can be done in sublinear time. If the vicinities of $u$ and $v$ intersect, the exact distance is returned. If not, a low stretch path can be retrieved by concatenating the paths from $u$ to $\ell(u)$, from $\ell(u)$ to $\ell(v)$, and finally $\ell(v)$ to $v$. This scheme can be generalized to further reduce space at the expense of increased stretch: rather than storing shortest paths between landmarks, we approximate these distances with the schemes of [15].

# 5 Average-degree Bounded Graphs are no harder than Maximum-degree Bounded Graphs

Our distance oracles achieve improved space/stretch trade-off with the minimalistic assumption on graph sparsity, that is, the total number of edges in the graph. One of the challenges in designing distance oracles with such minimalistic assumption is to handle skewed degree distribution of nodes in the graph. In this section, we show that in the context of designing distance oracles, average-degree-bounded graphs are no harder than maximum-degree-bounded graphs.

In particular, assume that we have a distance oracle $\mathscr{D}$ that is of size $O(S)$ and returns stretch-$s$ paths in $O(T)$ time for any $\Delta$-*degree bounded* graph on $n$ nodes, where $S$ and $T$ are functions of $n$, $\Delta$ and $s$. For any fixed stretch $s$ and fixed $\Delta$, we require $S(O(n)) = O(S(n))$ and $T(O(n)) = O(T(n))$, which is true for all functions $S$ and $T$ of interest since $S = O(n^2)$ and $T = O(n^2)$ for any non-trivial distance oracle. We show that $\mathscr{D}$ can be used to build a distance oracle of size $O(S)$ that returns stretch-$s$ paths on a graph with *average degree* $\Delta$ in at most $O(T)$ time.

Let $G = (V, E)$ be a connected graph with average degree $\Delta$. Given $G$, we will first create a $\Delta$-degree bounded graph $G_\Delta = (V_\Delta, E_\Delta)$. Then, we show how $\mathscr{D}$ can be used on $G_\Delta$ to return stretch-$s$ paths on $G$.

[2]The conference version of the paper [1] used the idea of vicinity-vicinity intersection, but we show in this version that ball-vicinity intersection suffices to achieve the same bounds.

**The Reduction.** For each node $v \in V$, create $\alpha_v = \lceil \deg(v)/\Delta \rceil$ nodes $v_1, v_2, \ldots, v_{\alpha_v}$ in $V_\Delta$. For each edge $e = (u, v) \in E$, if $\deg(u) \leq \Delta$ and $\deg(v) \leq \Delta$, create an edge $e = (u_1, v_1)$ in $E_\Delta$. For each node $v \in V$, we arbitrarily distribute $N(v)$ in $G$ to the nodes corresponding to $v$ in $G_\Delta$ such that for $i = 1, 2, \ldots, (\alpha_v - 1)$, $|N(v) \cap N(v_i)| = \Delta$ and $|N(v) \cap N(v_{\alpha_v})| = (\deg(v) - (\alpha_v - 1) \cdot \Delta)$. Finally, for each pair $v_i, v_{i+1}$, we create an edge in $E_\Delta$ of weight 0.

In order to answer an approximate distance query for any pair of nodes $u, v \in V$, we use $\mathscr{D}$ to answer approximate distance queries between $u_1, v_1 \in V_\Delta$ in $G_\Delta$ and let the length of the path returned by the data structure be $\delta'$. We output the distance $\delta'$ as an approximate distance for the pair of nodes in $G$.

**State and Query Time.** We first prove that asymptotically, the size of the data structure and the query time are not increased due to the reduction. Fix some stretch $s$. Recall that $S(O(n)) = O(S(n))$ and $T(O(n)) = O(T(n))$; all we need to show is that the number of nodes in $G_\Delta$ are within a constant factor of the number of nodes in $G$.

**Claim 1.** *$G_\Delta$ is a $\Delta$-degree bounded graph with $O(n)$ nodes.*

**Proof:** The degree boundedness is trivial from the construction. We prove the claim regarding number of nodes. The reduction implies that $|V_\Delta| = \sum_{v \in V} \lceil \deg(v)/\Delta \rceil = |V| + \sum_{v \in V} \lfloor \deg(v)/\Delta \rfloor$. This gives us an upper bound: $|V_\Delta| \leq |V| + \sum_{v \in V} \deg(v)/\Delta = 2 \cdot |V|$. □

**Stretch.** Consider any pair of nodes $u, v \in V$ at distance $d$ in $G$. It is trivial that in $G_\Delta$, the distance between the nodes $u_1, v_1 \in V_\Delta$ is $d' = d$. The data structure $\mathscr{D}$ returns a path of distance at most $\delta' = s \cdot d' = s \cdot d$, which is of stretch s.

In §8, we discuss how this reduction can be intuitively interpreted when it is incorporated into the algorithm of the next section, so that the algorithm runs directly on $G$ rather than $G_\Delta$. In the rest of the paper, we restrict our attention to $\Delta$-degree bounded graphs only.

# 6 Distance oracle for stretch 2

In this section, we present our distance oracles that return distances and paths of worst-case stretch 2. Throughout the section, we assume that the graph is a $\Delta$-degree bounded graph; the discussion in §5 then immediately gives us distance oracles for graphs with average degree $\Delta$. For any fixed $1 \leq \alpha \leq n$, our first distance oracle has size $O(n\Delta + n^2/\alpha)$ and returns stretch-2 distances in $O(\alpha\Delta)$ time. In §6.4, we show how to further reduce the query time to $O(\alpha)$ using an additional $O(n\Delta\alpha)$ space. Our oracles, similar to the oracles in [9,15], also allow retrieving paths in constant time per hop.

## 6.1 Constructing the distance oracle

Let $G = (V, E)$ be a $\Delta$-degree bounded graph. The construction begins by sampling each node independently at random with probability $1/\alpha$, creating a set $L$ of sampled "landmark" nodes. The distance oracle stores:

- For each node $v \in V$, a hash table containing its neighbors $N(v)$.
- For each node $v \in L$, a hash table containing the shortest distance to every other node in $G$.
- For each node $v \in V \backslash L$, $\ell(v)$ and the "ball radius" $r_v = d(v, \ell(v))$.

This completes the preprocessing of the graph and construction of the distance oracle. We start by proving the bound on the size of the distance oracle. Note that our construction of the distance oracle is randomized; the construction gives us a distance oracle that has size $O(n\Delta + n^2/\alpha)$ in expectation. Using a Chernoff bound, all our results hold with high probability with a logarithmic factor larger size. However, as discussed below, our query algorithm is deterministic; that is, it never outputs distances with stretch more than 2. Hence, a distance oracle with worst-case size $O(n\Delta + n^2/\alpha)$ can be constructed using a Las Vegas algorithm; see [15] for details.

**Claim 2.** *The size of the distance oracle is $O(n\Delta + n^2/\alpha)$, in expectation.*

**Proof:** Storing the list of neighbors for each node requires space $O(n\Delta)$. Note that $E[|L|] = n/\alpha$, and hence, storing shortest distances from nodes in $L$ to all nodes in the graph requires $O(n^2/\alpha)$ space in expectation. Storing $\ell(v)$ and $r_v$ requires $O(1)$ space for each node in $V \backslash L$. Hence, the total expected size is $O(n\Delta + n^2/\alpha + n) = O(n\Delta + n^2/\alpha)$. □

## 6.2 Answering distance queries

The algorithm Query-2$(u, v)$ to approximate the distance between nodes $u$ and $v$ is shown in Algorithm 1. Suppose the query asks for distance between nodes $u$ and $v$. The algorithms starts by running a shortest path algorithm that stops once $u$ and $v$ have computed their vicinities; such an algorithm, a modified version of the one presented in [15] for instance, takes time $O(\alpha\Delta)$. This can be done since the graph is stored in the distance oracle (in the form of neighbors for each node). Both $u$ and $v$ store their vicinities and the distances to each node in their vicinity in a hash table temporarily.

The algorithm then checks whether $v \in \Gamma(u)$ or $u \in \Gamma(v)$, in which case it directly reads $d(u, v)$ from the hash table maintained at $u$ or $v$ respectively. If $v \notin \Gamma(u)$ and $u \notin \Gamma(v)$, the algorithm performs a *ball-vicinity intersection check*: it queries each of the nodes $w \in B(u)$ and checks if $w \in \Gamma(v)$. If at least one such $w$ is found, it returns the minimum of $d(u, w) + d(w, v)$ over all such $w$. If there is no such $w$, the algorithm queries $u$ and $v$ for their ball radii $r_u$ and $r_v$. If $r_u < r_v$, the algorithm returns $d(u, \ell(u)) + d(\ell(u), v)$; else it returns $d(v, \ell(v)) + d(\ell(v), u)$.

**Algorithm 1** Query-2(u, v) – answering approximate distance queries with stretch-2.

```
Compute Γ(u), Γ(v)
If v ∈ Γ(u) or u ∈ Γ(v)
      return d(u, v)
Else if B(u) ∩ Γ(v) ≠ ∅
      return min_{w∈B(u)∩Γ(v)} {d(u, w) + d(v, w)}
Else if r_u ≤ r_v
      return d(u, ℓ(u)) + d(ℓ(u), v)
Else
      return d(v, ℓ(v)) + d(ℓ(v), u)
```

## 6.3 Analysis

We now analyze the query time and stretch for the above distance oracle and the query algorithm.

**Query time.** We start by analyzing the worst-case query time for the query algorithm. Using a standard argument as in [15, Lemma 3.2], for any node $v$ we have that $E[|B(v)|] = O(\alpha)$. Since the graph is $\Delta$-degree bounded for some fixed constant $\Delta$, we have that the vicinity size is bounded in expectation as follows: $E[|N(B(v))|] = \Delta \cdot E[|B(v)|] = O(\alpha\Delta)$.

To start with, the query algorithm requires constructing hash tables containing vicinities of the source and the destination. We note that computing the balls for the source and the destination using any of the standard shortest path algorithms takes time $O(\alpha\Delta)$ for a $\Delta$-degree bounded graph. Since the neighbors of each node are stored in the distance oracle, creating a hash table containing the vicinity takes an additional $O(\alpha\Delta)$ time. In the next step, the query algorithm checks for the ball-vicinity intersection; this takes an additional $O(\alpha)$ time – for each element in the ball of $u$, it takes $O(1)$ time to check whether it is contained in the hash table containing the vicinity of $v$. Hence, the total query time is bounded by $O(\alpha\Delta)$.

**Stretch** We obtain an upper bound of 2 on the stretch of the distance between the nodes returned by QUERY-2$(u, v)$. The proof uses the following lemma.

**Lemma 1** (**Ball-vicinity intersection lemma**). *For any pair of nodes $u, v \in V$, if $B(u) \cap \Gamma(v) = \emptyset$, the distance between $u$ and $v$ is lower bounded as $d(u, v) \geq r_u + r_v$.*

**Proof:** Let $B(u) \cap \Gamma(v) = \emptyset$. Let $P = (u, v_1, v_2, \ldots, v_k, v)$ denote the shortest path between $u$ and $v$ in $G$. Let $i_0 = \min\{i : v_i \in \Gamma(v)\}$. Since $B(u) \cap \Gamma(v) = \emptyset$, we have that $v_{i_0} \notin B(u)$. Hence, $d(u, v_{i_0}) \geq r_u$. Furthermore, note that $v_{i_0} \notin B(v)$ since otherwise $v_{i_0-1} \in \Gamma(v)$ which contradicts the definition of $i_0$. Hence, $d(v, v_{i_0}) \geq r_v$. Since $v_{i_0}$ lies on the shortest path along between $u$ and $v$, we get that $d(u, v) = d(u, i_0) + d(v, i_0) \geq r_u + r_v$. □

The above lemma shows that if two nodes $u$ and $v$ are "close", there must exist some node which lies in the vicinity of both $u$ and $v$. A word of caution though – the above lemma has to be interpreted in the right way – two nodes having a ball-vicinity intersection can be significantly far away. This is because the nodes in the vicinity of any node $u$ are not necessarily the $O(\alpha\Delta)$ closest nodes of $u$; while balls possess this structure, vicinities contain the neighbors of the nodes in the balls, thereby destroying any meaningful "distance based" interpretation of the vicinities. We now prove the bound on stretch.

**Claim 3.** *For any pair of nodes $u, v \in V$, if $d(u, v) < r_u + r_v$,* QUERY-2$(u, v)$ *returns the exact distance between $u$ and $v$.*

**Proof:** If $d(u, v) < r_u + r_v$, we have using Lemma 1 that there must be at least one node $x \in B(u)$, such that $x \in B(u) \cap \Gamma(v)$. The algorithm reads the "exact" distance $d(u, x)$ from the hash-table maintained at node $u$ and the "exact" distance $d(v, x)$ from the hash-table maintained at node $v$. From the proof of the ball-vicinity intersection lemma, we note that among all such nodes $x \in B(u) \cap \Gamma(v)$, there must be at least one node which lies on the shortest path between $u$ and $v$, and this node minimizes the distance returned by the algorithm resulting in stretch 1. □

For the case when $d(u, v) \geq r_u + r_v$, we show that our scheme results in a stretch at most 2.

**Theorem 1.** *For any pair of nodes $u, v \in V$, if $d(u, v) \geq r_u + r_v$, the algorithm* QUERY-2$(u, v)$ *returns a distance estimate of stretch at most* 2.

**Proof:** Without loss of generality, assume that $r_u \leq r_v$. Then, the condition in the lemma implies that $d(u, v) \geq 2 \cdot r_u$.

When $d(u, v) \geq r_u + r_v$, there are two possible cases. First, if $\exists x \in B(u) \cap \Gamma(v)$, the returned distance is $d(u, x) + d(v, x)$ for some $x \in B(u)$; or second, if $B(u) \cap \Gamma(v) = \emptyset$, the distance returned is $d(u, \ell(u)) + d(\ell(u), v)$. Let the distance returned by the query algorithm be $\delta(u, v) = d(u, w) + d(w, v)$ where either $w = \ell(u)$ or $w = \arg\min_{x \in B(u) \cap \Gamma(v)} \{d(u, w) + d(v, w)\}$. Note that $d(u, w) \leq d(u, \ell(u)) = r_u$.

By the triangle inequality, we have that $d(w, v) \leq d(w, u) + d(u, v)$. Hence, $\delta(u, v) \leq 2 \cdot d(u, w) + d(u, v)$. Since $d(u, w) \leq r_u$, we get $\delta(u, v) \leq 2 \cdot r_u + d(u, v)$. Using the lower bound of $2 \cdot r_u$ on the distance between $u$ and $v$, we get the desired bound of 2 on stretch. □

## 6.4 Storage versus computation

The distance oracle presented above allows one to smoothly trade-off query time to reduce the size of the distance oracle (by varying $\alpha$). In particular, this gives us distance oracles that require space linear in the size of the graph (by setting $\alpha = n^2/m$), which may be of independent interest. Moreover, for specific values of $\alpha$, one can also avoid running shortest path algorithm by using no additional space. More specifically, for $1 \leq \alpha \leq \sqrt{n}$, one can store the balls of each node within the distance oracle at no additional cost. For these values of $\alpha$, one can also design efficient compact routing schemes; we present such schemes in §9.

However, for other values of $\alpha$ and/or for even slightly dense graphs, there are two potential issues with the above distance oracle. First, the query algorithm requires running a (restricted) shortest path algorithm for each distance query; and second, for some fixed space, the query time can be high. For instance, for graphs with $\Delta = \Theta(n^{1/4})$, the distance oracle with $O(n^{7/4})$ space requires $O(n^{1/2})$ query time per query for stretch 2 distances.

In this subsection, we show how to reduce the query time for our distance oracle; for specific values of $\Delta$ and $\alpha$, this may increase the size of the distance oracle. In particular, we show how to reduce the query time to $O(\alpha)$ (reduce it by a factor of $\Delta$ when compared to the distance oracle above) using an additional $O(n\alpha\Delta)$ space. For the instance discussed above, for example, we will still be able to retrieve stretch 2 distances using $O(n^{7/4})$ space, using only $O(n^{1/4})$ query time.

In order to reduce the query time, we note that the above distance oracle requires computing vicinities for the nodes on the fly, leading to $O(\alpha\Delta)$ query time. Hence, in our new distance oracle, we preprocess the graph and store the vicinities within the distance oracle. Storing the vicinities, however, require $O(\alpha\Delta)$ space corresponding to each node and hence, the size of the new distance oracle is $O(n\alpha\Delta + n^2/\alpha)$. The reduction in query times comes due to the fact that if the vicinities are already stored within the distance oracle, checking for ball-vicinity intersection takes $O(\alpha)$ time – for each element in the ball of $u$, it takes $O(1)$ time to check whether it is contained in the hash table containing the vicinity of $v$. Since we do not require running any shortest path algorithm, the query time of the algorithm is bounded by the time taken to check for ball-vicinity intersection and hence, is reduced to $O(\alpha)$.

We remark that the two distance oracles presented above may achieve different set of operating points within the space/query time trade-off. For example, as discussed above, for graphs with $\Delta = \Theta(n^{1/4})$, it is not possible to retrieve stretch 2 paths using $O(n^{7/4})$ space and $o(n^{1/2})$ query time using the first distance oracles; the second distance oracle does achieve these new operating points. On the other hand, it is not possible to achieve linear space using the second distance oracle. Nevertheless, the second distance oracle – besides having a lower query time – has the additional benefit that the query algorithm does not run shortest path algorithm for each query.

## 6.5 Discussion

**Implications of the average-to-max-degree-bound reduction** The results in this section combined with the reduction of §5 immediately give us distance oracles of size $O(n\Delta + n^2/\alpha)$ and $O(n\Delta\alpha + n^2/\alpha)$, which for any graph with at most $O(n\Delta)$ edges, returns stretch-2 paths in $O(\alpha\Delta)$ and $O(\alpha)$ time, respectively. We show how to incorporate the reduction into the algorithm in a way that yields intuition and eases implementation.

Specifically, let $G$ be the graph with average degree $\Delta$. The reduction implies that each node $v$ in $G$ which has degree $\deg(v) > \Delta$ effectively "emulates" $\lceil \deg(v)/\Delta \rceil$ nodes in $G_\Delta$. Now consider constructing the distance oracle presented in this section. While sampling nodes for the landmark set $L$, the node $v$ is now sampled with probability $1/\alpha \cdot \lceil \deg(v)/\Delta \rceil$, that is, with probability that is proportional to the degree of $v$. Moreover, due to Claim 1, the size of $B(v)$ remains unchanged asymptotically ($B(v)$ and hence $\Gamma(v)$ may change, but not their size). Thus, the implications of the reduction are simple: just sample each node $v$ in the graph with probability $1/\alpha \cdot \lceil \deg(v)/\Delta \rceil$ rather than probability $1/\alpha$. *In other words, rather than sampling nodes uniform-randomly, they are sampled with probability proportional to their degree.*

**An optimization** Although the worst-case stretch for our distance oracle is 2, we can apply simple heuristics to improve the stretch in practice. Recall that the worst-case stretch in our distance oracle occurs for source-destination pairs $u, v$ for which $B(u) \cap \Gamma(v) = \emptyset$; the query may return a path, for instance, $u \rightsquigarrow \ell(u) \rightsquigarrow v$ that is of stretch 2. The main observation is that for such source-destination pairs, there may exist a $w \in \Gamma(u)$ for which the length of the path $u \rightsquigarrow w \rightsquigarrow \ell(w) \rightsquigarrow v$ is less than the path $u \rightsquigarrow \ell(u) \rightsquigarrow v$. The approximate distance query can then be answered by the distance oracle as the minimum of the distances retrieved by checking all $w \in \Gamma(u)$ (see §9 for implementation details). Since checking the length of the paths $u \rightsquigarrow w \rightsquigarrow \ell(w) \rightsquigarrow v$ for all $w \in \Gamma(u)$ takes (asymptotically) the same time as checking the ball-vicinity intersection, the heuristic does not increase the query time, with potential improvements in stretch of retrieved paths. Indeed, we show in §10 that this heuristic increases the number of source-destination pairs that retrieve shortest paths by almost 25%.

# 7 Distance oracles for stretch 3 and larger

In this section, we present distance oracles that return paths of worst-case stretch $(4k-1)$, for any positive integer $k$. For any fixed $1 \leq \alpha \leq n$, the distance oracle is of size $O(n\Delta + (n/\alpha)^{(1+1/k)})$ and returns stretch-$(4k-1)$ distances in $O(\alpha\Delta)$ time for any graph with $O(n\Delta)$ edges; the paths can then be retrieved in constant time per hop. In particular, we get a distance oracle of size $O(n\Delta + n^2/\alpha^2)$ that returns stretch 3 distances in $O(\alpha\Delta)$ time. As earlier, the query time can be further reduced to $O(\alpha)$ using an extra $O(n\alpha\Delta)$ space and while we present bounds that hold in expectation, our distance oracles can be derandomized using a Las Vegas algorithm.

## 7.1 Constructing the distance oracle

Let $G = (V, E)$ be a $\Delta$-degree bounded graph. Fix some $1 \leq \alpha \leq n$ and some integer $k > 0$. Our construction of distance oracle $\mathscr{D}$ begins by sampling each node independently at random with probability $1/\alpha$, creating a set $L$ of sampled nodes. We now create a complete graph $G'$ with nodes in $L$ as the node set and for each pair $l_1, l_2 \in L$, the weight of the edge $(l_1, l_2)$ being the shortest path between $l_1$ and $l_2$ in $G$. We run the approximate distance oracle from Thorup and Zwick on $G'$ to construct a distance oracle $\mathscr{D}'$ that stores $(2k-1)$-approximate shortest paths between every pair of nodes in $L$. $\mathscr{D}$ stores $\mathscr{D}'$ as a sub-data structure. Furthermore, $\mathscr{D}$ also stores, for each node $v \in V$, its set of neighbors $N(v)$, its closest landmark node $\ell(v)$ and the ball radius $r_v$.

**Claim 4.** *The size of the distance oracle is* $O\left(n\Delta + (n/\alpha)^{(1+1/k)}\right)$, *in expectation.*

**Proof:** Note that $E[|L|] = O(n/\alpha)$ and hence, using the results in [15], we get that the size of the distance oracle $\mathscr{D}'$ is $O((n/\alpha)^{(1+1/k)})$. Furthermore, storing $N(v)$ for each node $v$ requires an additional $O(n\Delta)$ space; storing $\ell(v)$ and $r_v$ require an additional $O(1)$ space. Hence, the size of the distance oracle is $O(n\Delta + (n/\alpha)^{(1+1/k)})$, in expectation. □

## 7.2 Answering distance queries

Let QueryTZ(u, v) be the query algorithm for the Thorup-Zwick scheme [15] that returns $(2k-1)$-approximate distances between nodes $u$ and $v$. The query algorithm for our distance oracle is shown in Algorithm 2.

Suppose the query asks for distance between nodes $u, v \in V$. The algorithm, as in Algorithm 1, starts by running a shortest path algorithm that stops when once the two nodes $u$ and $v$ have computed their vicinities and shortest distances to nodes in their vicinities. This can be done since the graph is stored in the distance oracle (in the form of an adjacency list) and requires $O(\alpha\Delta)$ time using a modified version of the algorithm presented in [15]. Both $u$ and $v$ temporarily store this information in a hash table.

If $v \in \Gamma(u)$ or $u \in \Gamma(v)$, the algorithm returns the exact distance $d(u,v)$ from the hash table at $u$ or $v$, respectively. If $v \notin \Gamma(u)$ and $u \notin \Gamma(v)$, the algorithm checks for *ball-vicinity intersection*, that is, for each node $w \in B(u)$, the algorithm checks if $w \in \Gamma(v)$. If at least one such $w$ is found, the algorithm returns the minimum of $d(u,w) + d(v,w)$ over all such $w$. If no such $w$ exists, the algorithm returns $d(u,\ell(u)) + \text{QueryTZ}(u,v) + d(v,\ell(v))$. Finally, the hash tables are deleted from nodes $u$ and $v$.

**Algorithm 2** Query$(u,v)$ – answering approximate distance queries with stretch-$(4k-1)$. QueryTZ(u, v) is the query algorithm for the Thorup-Zwick scheme that, for any integer $k > 0$, returns $(2k-1)$-approximate distances between nodes $u$ and $v$.

```
Compute Γ(u), Γ(v)
If v ∈ Γ(u) or u ∈ Γ(v)
      return d(u,v)
Else if B(u) ∩ Γ(v) ≠ ∅
      return min_{w∈B(u)∩Γ(v)} {d(u,w) + d(v,w)}
Else
      return d(u,ℓ(u)) + QueryTZ(u,v) + d(ℓ(v),v)
```

## 7.3 Analysis

In terms of the query time, we note that the above query algorithms is very similar to the query algorithm for our distance oracles with stretch 2; indeed, the only difference is the final step of the query algorithm, that is, when $B(u) \cap \Gamma(v) = \emptyset$. Since checking ball-vicinity intersection is the bottleneck in terms of query time, we get, using arguments similar to those in §6, that the query time for the above query algorithm is $O(\alpha\Delta)$. Moreover, since the definition of balls and vicinities for the above distance oracle are exactly the same as those in §6, using exactly the same proofs as in §6, we get the following claims:

**Claim 5.** *For any pair of nodes u and v, if $B(u) \cap \Gamma(v) = \emptyset$, the distance between u and v is lower bounded as $d(u,v) \geq r_u + r_v$.*

**Claim 6.** *For any pair of nodes u and v, if $d(u,v) < r_u + r_v$,* Query$(u,v)$ *returns the exact distance between u and v.*

When $d(u,v) \geq r_u + r_v$, we have two cases similar to those in the proof of Theorem 1. For the first case, when $B(u) \cap \Gamma(v) \neq \emptyset$, we get the following claim the proof of which follows from the proof of Theorem 1:

**Claim 7.** *For any pair of nodes u and v, if $d(u,v) \geq r_u + r_v$ and $B(u) \cap \Gamma(v) \neq \emptyset$,* Query$(u,v)$ *returns a distance of stretch at most* 2 *between u and v.*

The only remaining case is when $d(u,v) \geq r_u + r_v$ and $B(u) \cap \Gamma(v) = \emptyset$; for this case, we prove a worst-case stretch bound of $(4k-1)$:

**Theorem 2.** *If $d(u,v) \geq r_u + r_v$ and $B(u) \cap \Gamma(v) = \emptyset$, the algorithm* Query$(u,v)$ *returns, in the worst case, distance estimate of stretch-$(4k-1)$ between u and v.*

**Proof:** When $d(u,v) \geq r_u + r_v$ and $B(u) \cap \Gamma(v) = \emptyset$, the distance returned by the scheme is $\delta(u,v) = d(u,\ell(u)) + \text{QueryTZ}(u,v) + d(\ell(v),v)$. Since QueryTZ(u, v) returns $(2k-1)$-approximate distances, we have that $\delta(u,v) \leq d(u,\ell(u)) + (2k-1)d(\ell(u),\ell(v)) + d(\ell(v),v)$. By the triangle inequality, $d(\ell(u),\ell(v)) \leq d(\ell(u),u) + d(u,v) + d(v,\ell(v))$. Hence, $\delta(u,v) \leq 2k \cdot d(u,\ell(u)) + (2k-1)d(u,v) + 2k \cdot d(v,\ell(v))$. Since $d(u,\ell(u)) = r_u$ and $d(u,\ell(v)) = r_v$, we get $\delta(u,v) \leq 2k \cdot r_u + (2k-1)d(u,v) + 2k \cdot r_v = 2k(r_u + r_v) + (2k-1)d(u,v)$. Using the condition of the theorem, we get $\delta(u,v) \leq 2k \cdot d(u,v) + (2k-1)d(u,v) = (4k-1) \cdot d(u,v)$, which we set out to prove. □

### 7.4 Discussion

We close the section with remarks along the lines of §6.4 and §6.5. First, using ideas similar to those in §6.4, the query time for the above distance oracle can be reduced to $O(\alpha)$ by storing the vicinities within the distance oracle – this requires an additional $O(n\alpha\Delta)$ space. This allows us to achieve new points within the space/query time trade-off although it is no more possible to have size linear in the size of the graph.

Next, although we assumed that the input graph is $\Delta$-degree bounded, results from §5 imply that the results generalize to graphs with average degree $\Delta$. As in §6.5, we only require to sample nodes for inclusion in the landmark set with a probability proportional to the degree of the node rather than sampling them uniform randomly.

Finally, an optimization similar to that in §6.5 is again possible for the algorithm described in this section. While checking for ball-vicinity intersection, when $u$ queries each of the nodes $w \in B(u)$, it could actually query for its distance to $\ell(w)$ and combine this with $d(\ell(w), \ell(v))$ and $d(v, \ell(v))$ for a potentially better estimate of the distance between $u$ and $v$. Again, the approximate distance query can then be answered by the distance oracle as the minimum of all the distances retrieved by querying the nodes in the ball of the source resulting in improved stretch in practice without any asymptotic increase in the query time.

## 8 Distance oracles with additive stretch

In this section, we show that the space/query time trade-off in our distance oracles from §6 and §7 can be further improved at the cost of a small additive stretch. In particular, let $G$ be an unweighted graph and let $u, v$ be a pair of nodes at distance $d$; then, for any fixed $1 \leq \alpha \leq n$, we design:

- a distance oracle of size $O(n\alpha + n^2/\alpha)$ that returns distances of at most $2d + 1$ in time $O(\alpha)$, and
- a distance oracle of size $O\left(n\alpha + (n/\alpha)^{(1+1/k)}\right)$ that returns distances of at most $(4k-1)d+2k$ in time $O(\alpha)$

The results can be generalized to weighted graphs without any increase in space or query time. As earlier, while the bounds presented hold in expectation, our construction algorithm can be derandomized using a Las Vegas algorithm.

The main observation that allows us to design these distance oracles is captured in the following lemma, which presents a lower bound on distance between the source and the destination when the query algorithm, rather than checking for ball-vicinity intersection as in Lemma 1, checks only for ball-ball intersection:

**Lemma 2 (Ball-ball intersection).** *For any pair of nodes $u, v \in V$, let $w_{uv}$ be the weight of the heaviest edge along the shortest path between $u$ and $v$. If $B(u) \cap B(v) = \emptyset$, the distance between $u$ and $v$ is lower bounded as $d(u, v) \geq r_u + r_v - w_{uv}$.*

**Proof:** Assume that $B(u) \cap B(v) = \emptyset$ and let $P = (u, x_1, x_2, \ldots, v)$ be the shortest path between $u$ and $v$. Let $i_0 = \max\{i | x_i \in P \cap B(u)\}$, $w = x_{i_0}$ and $w' = x_{i_0+1}$. By definition, $w' \notin B(u)$ and hence, $d(u, w') \geq r_u$. Furthermore, since $w$ and $w'$ are neighbors and $w \in B(u)$, we have that $d(u, w) \geq r_u - w_{uv}$. Furthermore, since $B(u) \cap B(v) = \emptyset$, $w \notin B(v)$ leading to the fact that $d(v, w) \geq r_v$; since $w$ is on the shortest path between $u$ and $v$, we have that $d(u, v) = d(u, w) + d(v, w) \geq r_u + r_v - w_{uv}$. □

Lemma 2 suggests that if the query algorithms from §6 and §7 were to perform ball-ball intersection check rather than ball-vicinity intersection check, the loss in stretch can be bounded by a constant factor that depends on the heaviest weight along the shortest path between the source and the destination. In contrast to the distance oracles of §6 and §7, performing ball-ball intersection neither requires storing the vicinities nor computing them on the fly; query is now performed only on the balls of each node leading to improvements in space and/or query time.

## 8.1 Constructing the distance oracles

Let $G = (V, E)$ be a $\Delta$-degree bounded graph. The construction begins by sampling each node independently at random with probability $1/\alpha$, creating a set $L$ of sampled "landmark" nodes.

**Distance oracle for additive stretch 1.** The distance oracle is similar to the one in §6.4. It stores, for each node $v \in L$, a hash table containing the shortest distance to every other node in $G$ and for each node $v \in V \setminus L$, distances to nodes in its ball, its landmark node $\ell(v)$ and the "ball radius" $r_v = d(v, \ell(v))$.

To bound the size of the distance oracle, we note that we have $O(n/\alpha)$ landmarks, in expectation, requiring $O(n^2/\alpha)$ space to store distances to each other node in the graph. Furthermore, each node has $O(\alpha)$ nodes in its ball and hence, storing distances to these nodes require $O(n\alpha)$ space; storing $\ell(v)$ and $r_v$ for each node $v$ requires an additional $O(1)$ space. Hence, the total space requirements are $O(n\alpha + n^2/\alpha)$, in expectation.

**Distance oracle for additive stretch $2k$.** The distance oracle is again very similar to the one in §7. First, a complete graph on nodes in $L$ is computed, where weight of each edge is equal to the shortest distance between the two nodes. The distance oracle $\mathscr{D}$ stores, as a sub-data structure, the Thorup-Zwick distance oracle $\mathscr{D}'$ that returns stretch $(2k-1)$ distances for the complete graph over nodes in $L$. In addition, $\mathscr{D}$ stores, for each node $v \in V \setminus L$, distances to nodes in its ball, its landmark node $\ell(v)$ and the "ball radius" $r_v = d(v, \ell(v))$.

Recall that the expected number of nodes in the landmark set is $O(n/\alpha)$ and hence, size of the sub-data structure $\mathscr{D}'$ is $O((n/\alpha)^{(1+1/k)})$. Furthermore, since the size of ball for each node is $O(\alpha)$, the additional space required is $O(\alpha)$ for each node. The overall size of the distance oracle is, hence, $O(n\alpha + (n/\alpha)^{(1+1/k)})$, in expectation.

## 8.2 Query algorithms and analysis

The query algorithms for the above distance oracles are similar to their respective query algorithms from §6 and §7 with the only change that it performs ball-ball intersection check rather than ball-vicinity intersection check. Regarding the query time, we note that since balls for each node are stored within the distance oracles, checking for ball-ball intersection requires $O(\alpha)$ time, leading to the claimed bound on the query time (all other operations require constant time).

We prove the stretch bound for the first distance oracle; for larger stretch, the proof follows using simple modifications.

**Theorem 3.** *For any two nodes $u, v \in V$ at distance $d$, let $w_{uv}$ be the weight of the heaviest edge along the shortest path between $u$ and $v$. Then, the query algorithm returns a distance of at most $2d + w_{uv}$.*

**Proof:** For the case when $d(u, v) < r_u + r_v - w_{uv}$, using Lemma 2, it is easy to show that the query algorithm returns the exact distance between $u$ and $v$.

Consider the case when $d(u, v) \geq r_u + r_v - w_{uv}$ and without loss of generality, assume that $r_u \leq r_v$. Then, the condition implies that $d(u, v) \geq 2 \cdot r_u - w_{uv}$. In such a case, the distance returned by the query algorithm is $d(u, \ell(u)) + d(\ell(u), v)$. By the triangle inequality, we have that $d(\ell(u), v) \leq d(\ell(u), u) + d(u, v)$. Hence, $\delta(u, v) \leq 2 \cdot d(u, \ell(u)) + d(u, v)$. Since $d(u, \ell(u)) \leq r_u$, we get $\delta(u, v) \leq 2 \cdot r_u + d(u, v)$. Using the lower bound of $2 \cdot r_u - w_{uv}$ on the distance between $u$ and $v$, we get the desired bound of $2d(u, v) + w_{uv}$ on stretch. □

Similarly, one can prove that for any pair of nodes $u, v$ at distance $d$, the second distance oracle returns a distance of at most $(4k-1)d + 2k \cdot w_{uv}$, where $w_{uv}$ is the weight of the heaviest weight along the shortest path between $u$ and $v$.

# 9 Compact routing schemes

Work on compact routing has applied the traditional results from approximate distance oracles [15] to network routing problems [14] in order to route the packets along short paths while using little memory at routers. These solutions have been proposed as centralized algorithms [14] and more recently as distributed protocols for wireless sensor networks [8], the Internet [11] and peer-to-peer networks [3]. In this section, we present compact routing schemes for our distance oracles; by exploiting graph sparsity, our schemes significantly improve the memory/stretch trade-off from previously known results. In particular, we discuss a surprisingly lightweight scheme that can be incorporated in distributed routing protocol implementations of the Thorup-Zwick (TZ) scheme, [11] for instance, to get a distributed routing protocol for our distance oracles.

We primarily focus on designing compact routing schemes for stretch 2. Recall that our distance oracle for stretch 2 has size $O(n\Delta\alpha+n^2/\alpha)$; our scheme distributes the state uniformly across all routers, requiring each router to store $O(\Delta\alpha + n/\alpha)$ entries. Using $\alpha = \sqrt{n/\Delta}$, our scheme requires each router to store $O(\sqrt{n\Delta})$ entries, while routing along paths of stretch 2. For graphs with $\Delta = o(n)$, this gives us the first scheme that routes along paths of stretch less than 3 and requires sublinear state at routers in the network. In fact, for real-world networks, that is networks with $\Delta = \Theta(\text{polylog}(n))$, our compact routing scheme requires the same amount of memory as [3,8,11,14] but routes along paths that have worst-case stretch bounded by 2. Note that any routing scheme with stretch less than 2 must require linear state at some node in the network [6] even for extremely sparse graphs; our scheme, hence, achieves the optimal stretch with non-trivial memory requirements at routers.

In addition, by setting $\alpha = \sqrt{n}$ in results from §8, we get a compact routing scheme that, for any source-destination pair at distance $d$, routes along paths of length at most $2d + 1$ by using $O(\sqrt{n})$ memory at each router – independent of the density of the graph.

**TZ scheme and our distance oracle.** Our distance oracle can be incorporated into the the proposed distributed adaptations [3,8,11] of the TZ scheme with minimal changes. This is due to the fact that the construction in our distance oracle, in concept, is similar to the TZ scheme: both schemes construct a set $L$ of nodes and each node $v$ stores a corresponding nearest neighbor $\ell(v)$ and certain nodes in its neighborhood. The first difference between our distance oracle and the TZ scheme is that the set $L$ is sampled proportional to node degree rather than uniform-randomly. Second, our distance oracle differs from TZ scheme in terms of the information stored in the distance oracle: for any node $v$, while TZ only requires storing the ball $B(v)$, our distance oracles stores $\Gamma(v)$. Both modifications are easy changes to the distributed protocols of [3,8,11]; note that computing $\Gamma(v)$ requires only neighbors of nodes in $B(v)$. Third, to route from the source $u$ to the destination $v$, our distance oracle allows $u$ to set up an initial connection to $v$ by using the TZ algorithm for routing between $u$ and $v$. This initial connection gives a path of stretch 3, via an essentially unmodified proof of [14,15]. The final task is to improve the stretch from 3 to 2.

**Implementing ball-vicinity intersection.** In order to improve the stretch from 3 to 2, our distance oracle requires the source and the destination to perform a ball-vicinity intersection (see Lemma 1). We show how vicinity intersection can be implemented in practice with a surprisingly lightweight *handshaking* scheme; that is, exchange of very few bytes between the source and the destination. Recall, from the discussion above, that the initial connection gives the source a path to the destination with stretch 3. The source can then send the list of nodes in its *ball* to the destination using this path. For the router-level map of the Internet measured by CAIDA [13], which consists of $n = 192,244$ routers and has average degree $\Delta \simeq 0.4 \log_2 n$, this requires the source to transfer roughly $4 \cdot \sqrt{n/\Delta}$ bytes, since IPv4 addresses are 4 bytes and balls have size $\sqrt{n/\Delta}$. This amount to approximately 661 bytes of data; on today's Internet, packets are generally allowed to be at least 1500 bytes long, so this would take just one packet.

The destination can then perform a ball-vicinity intersection, which requires $O(\sqrt{n/\Delta})$ time asymptotically but using the above numbers requires less than 165 hash table lookups which is fast in practice.[3] Upon executing the ball-vicinity intersection, the destination informs the source whether the ball-vicinity intersection is an empty set or not. If they do intersect, it can inform the source of the node (or nodes) at which ball-vicinity intersection occurs. This requires at most one packet which can be routed from the destination through the source via a stretch-3 path. The source-destination pair, after the above handshaking scheme (that requires at most two packets), now have a route with stretch 2.

In practice, this is likely to be efficient even for relatively short-lived connections. For much larger networks, of course, the exchange of ball information would require more bandwidth and computation; but since a stretch-3 path is available immediately, the reduction to stretch 2 can be treated as an optimization for longer flows in order to amortize the overhead.

**Probing and Shortcutting.** The protocol for implementing ball-vicinity intersection discussed above does not exploit the optimization discussed in §6.5 and §7.4 for heuristically improving the stretch for the retrieved paths. We discuss the implementation aspects related to the optimization. Implementing the optimization in practice leads to a process, which we call *probing and shortcutting* (P&S). P&S requires the source node to *probe* the nodes in its vicinity for improving stretch. We argue that this can be achieved with an extremely low overhead probing scheme. Once the source node finds a node in its vicinity that provides a better stretch, the source can conveniently switch the traffic through the *shortcut* path. We only discuss the probing mechanism, since shortcutting can be implemented easily in practice (note that the destination is oblivious to the shortcutting mechanism and hence, P&S does not require any handshaking mechanism).

For the probing mechanism, assume that the source opens an initial connection to a destination. The source, every $10^{\text{th}}$ packet, can probe a node in its vicinity (the question on deciding an appropriate order of probing the nodes in vicinity is discussed below) requesting the length of the path available from this node to the destination. These packets can be extremely small compared to the other data packets, leading to an extremely small overhead in terms of bandwidth consumed (just a fraction 0.1 more packets that are of negligible size compared to the data packets). Since the source-destination connections that account for most of the bandwidth sent on the networks are very long [16], we believe it is reasonable to amortize the cost of the probing over the lifetime of the connection.

In terms of the order of probing, we consider two heuristics. *Farthest-first*, in which the source probes the nodes that are the *boundary* nodes of its vicinity; and, *closest-first*, in which the source performs probing starting with the closest nodes (its neighbors). We show, through evaluations, that the former performs better than the latter.

# 10 Evaluation Results

In this section, we evaluate the performance of our stretch 2 and stretch 3 schemes on large-scale synthetic and realistic topologies. We first present our methodology, followed by a summary of the evaluation results and conclude with a detailed discussion on the results.

## 10.1 Methodology

**Schemes.** We evaluate three schemes: the stretch-3 scheme of Thorup and Zwick (TZ) [14,15]; **Reduced approximation ratio (ReAR)**: the stretch-2 scheme from §6 with $\alpha = \sqrt{n}$; and **Reduced space (ReS)**: the stretch-3 scheme (for $k = 1$) from §7 with $\alpha = \sqrt{n}$. Furthermore, we evaluate ReAR and ReS schemes with and without the P&S optimization discussed in earlier sections. For the TZ scheme, we sampled each node (for set $L$) with probability $\sqrt{\log n}/\alpha$. For ReAR and ReS, each node was sampled with probability $\sqrt{\log n}/\alpha \times \deg(v)/\log^2 n$. All the constants in the big-O notation were set to be 1. For our ReAR and ReS schemes, we implemented a modified scheme where we perform vicinity-vicinity intersection rather than ball-vicinity intersection; this requires slightly higher query time (an extra $\log n$ factor for the networks that we evaluated our schemes for) but may lead to improved stretch values.

[3] If the destination is a server, this could be a burden; but note that we could just as easily flip the protocol around so the source does the computation.

**Simulator.** We wrote a static simulator to simulate the above schemes. Hence, from the perspective of application to distributed compact routing protocols, the results presented in this section assume a static network topology and give post-convergence results only. As outlined in §9, a distributed implementation of our stretch-2 scheme is a straightforward extension of past work, but we leave a full dynamic evaluation to future work. Our static simulator allows us to evaluate the schemes at much larger scale.

**Topologies.** We present evaluation results for three topologies. (1) $G(n,m)$ random graphs, *i.e.*, $n = 16384$ nodes with $m$ uniform-random edges, with $m$ set so that the average degree is 6, (2) geometric random graphs with $n = 16384$ nodes with average degree 6, and (3) a 33,014 node AS-level map of the Internet (referred to as the Internet graph in this section) [13].

For $G(n,m)$ graphs and the Internet graph, link weights are 1; for geometric random graphs, a link's weight is the Euclidean distance between the position of its two nodes. For $G(n,m)$ graphs and for geometric random graphs, we generated 10 different topologies with the same parameters and our results are the average of evaluations of these topologies. For geometric random graphs, we sampled a set of "source" nodes and evaluated the performance of the schemes from these sources to all the destinations. We found that sampling 1/4 of the nodes as sources provided accurate results.

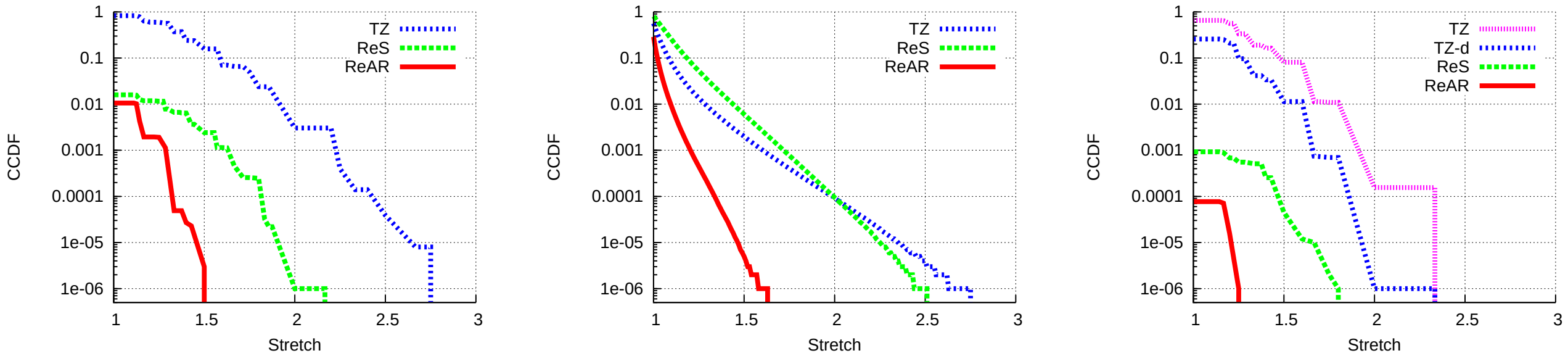

**Figure 1:** Complementary CDF of Stretch in $G(n,m)$ random graph (left), geometric random graph (middle) and Internet graph (right).

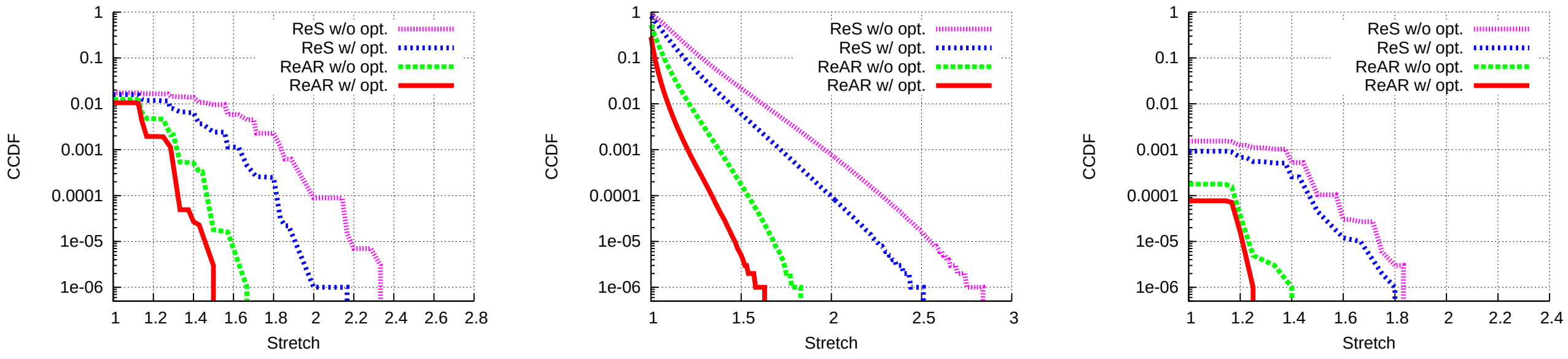

**Figure 2:** Complementary CDF of Stretch (ReAR and ReS) in $G(n,m)$ random graph (left), geometric random graph (middle) and Internet graph (right).

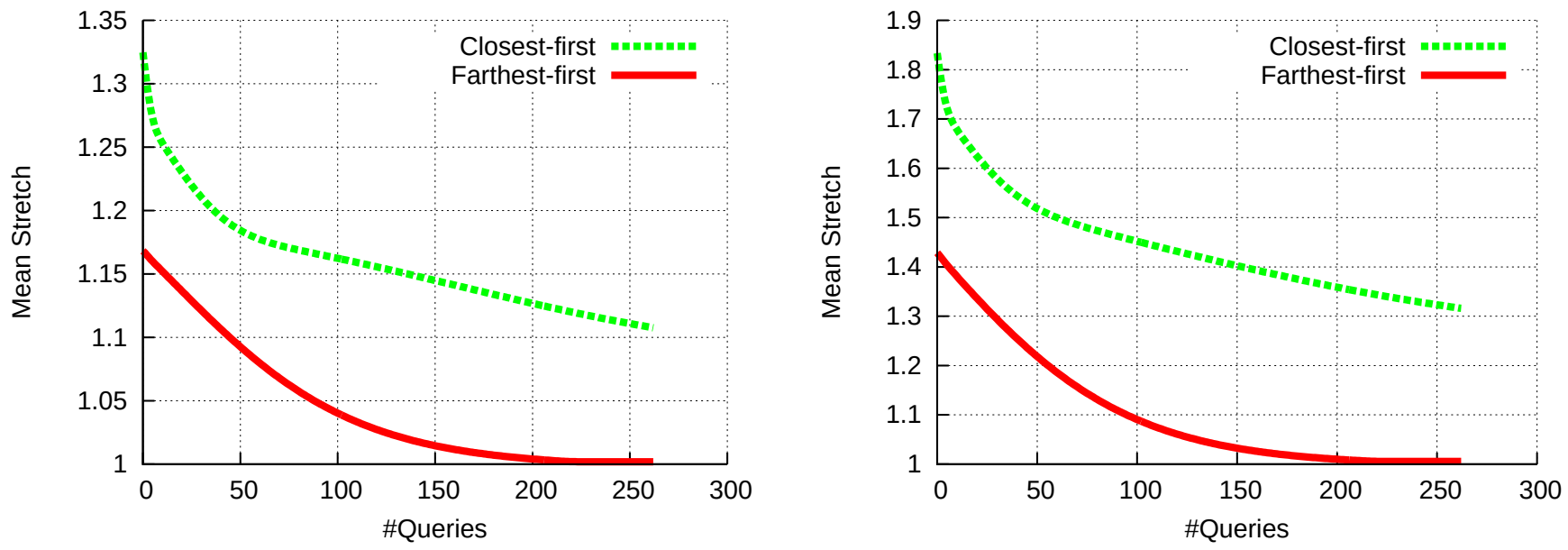

**Figure 3:** Mean stretch versus query time for ReAR (left) and ReS (right) for 16,384 node $G(n, m)$ graph with average degree 6.

## 10.2 Results and Discussions

**Stretch comparison with the TZ-scheme.** Fig. 1 shows the performance of the three schemes for various graph topologies (TZ is the original TZ scheme, TZ-d scheme is discussed below). The most notable result of this evaluation is that ReAR allows retrieval of exact shortest paths for nearly all source-destination pairs: 98.94% in the $G(n, m)$ graph, and 99.98% in the Internet graph. Though $G(n, m)$ graphs and the Internet graph have highly different structures, these graphs have a common feature: for nearly all source-destination pairs, the two vicinities intersect, thus providing a shortest path. In the $G(n, m)$ graph (in which 96.2% source-destination pairs have intersecting vicinities), this occurs since, with high probability, the diameter of the graph is roughly at most twice the vicinity radius. In the Internet graph (in which 96.8% source-destination pairs have intersecting vicinities), vicinity intersection likely occurs at the "core" networks of the Internet. Since TZ scheme does not exploit the vicinity intersection, its performance is significantly worse than our schemes (only 34.4% of the source-destination pairs retrieved shortest paths).

The surprising difference between the performance of the two schemes may be due to the difference in which these schemes construct the landmark set $L$. We evaluated a modified version of the TZ scheme that uses the same set $L$ as used by our schemes (see TZ-d in Fig. 1). Although this improves the performance of the TZ scheme (74.2% of the source-destination pairs now retrieve shortest paths), it is still much worse than the ReAR and ReS schemes. We, hence, believe that the high performance of our schemes is indeed due to the vicinity intersection idea.

For geometric random graphs, ReAR allows retrieval of shortest paths only for 70.7% of the source-destination pairs in comparison to 42.9% for the TZ scheme; indeed, only 4.8% of the source-destination pairs have intersecting vicinities. However, ReAR consistently performs better than the TZ-scheme, which in turn performs better than ReS. Finally, while the TZ-scheme performs better than ReS on an average, the worst-case stretch for the TZ-scheme is consistently worse than ReS. We believe that this is due to the P&S optimization, that allows many source-destination pairs to retrieve shorter paths due to short-cutting.

**Stretch comparison of ReAR and ReS.** The performance of ReAR and ReS for various graph topologies is compared in Fig. 2. We note that, as expected, ReAR consistently performs better than ReS, even without the P&S optimization. However, the more interesting observation is that the P&S optimization is much more effective in ReS. In particular, we note that the tail of ReS without the P&S optimization is significantly reduced when the optimization is used.

For $G(n, m)$ graphs, the stretch for 99% of the source-destination pairs is less than 1.15 using ReAR. For ReS, this is almost 1.3 (optimized version) and 1.5 (unoptimized version). The case of geometric random graphs is rather interesting: first, we observe that not many source-destination pairs have intersecting vicinities, otherwise ReS without the P&S optimization would not have achieved such a low fraction of source-destination pairs retrieving shortest paths (only around 11%). Despite this, ReAR performs surpris-ingly well: almost 48% of the source-destination pairs retrieve shortest paths without the P&S optimization and almost 71% retrieve shortest paths with the P&S optimization.

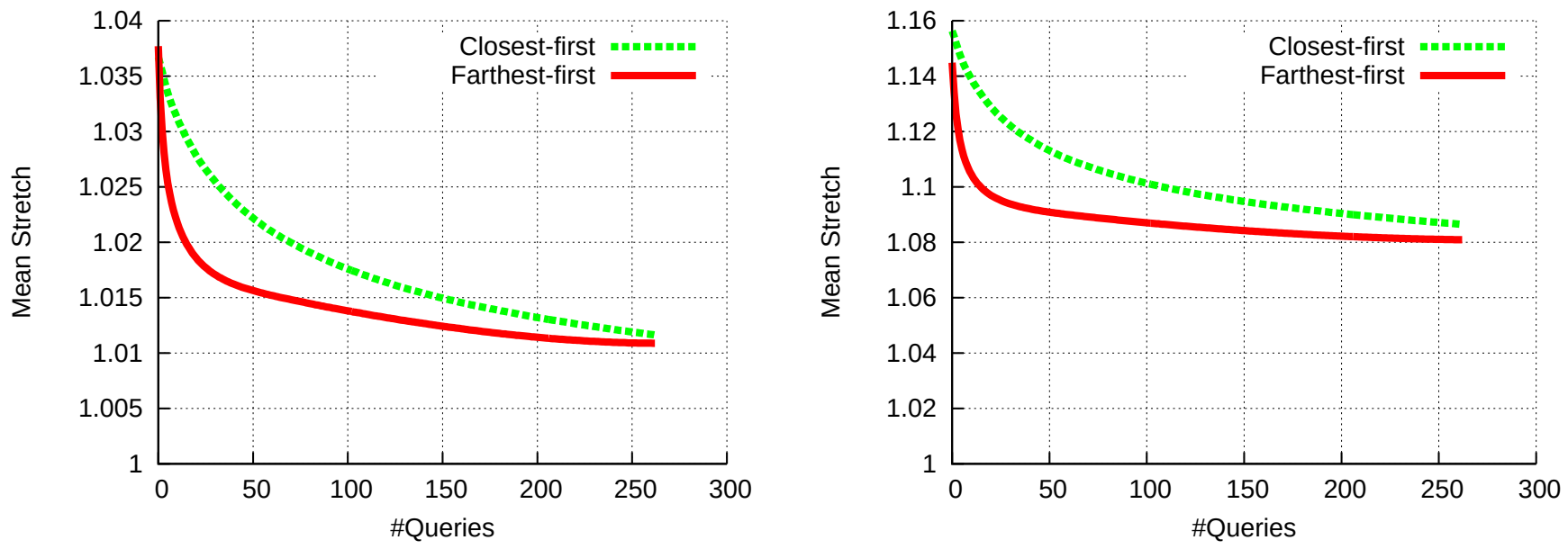

**Figure 4:** Mean stretch versus query time for ReAR (left) and ReS (right) for 16,384 node geometric random graph with average degree 6.

**Stretch versus Query Time.** For $G(n,m)$ graphs, Fig. 3 shows the variation of *mean stretch* – averaged over all source-destination pairs – with the number of queries for ReAR and ReS schemes, for the farthest-first and closest-first heuristics discussed in §9. We see a clear trend of "diminishing returns" where a few initial queries significantly reduce the stretch compared to no queries, after which the improvement is minimal. Based on the results, we conclude that in general, the farthest-first heuristic performs better in terms of the stretch with smaller query time. For the same two heuristics for stretch versus query time, Fig. 4 shows the results for ReAR and ReS schemes for the geometric random graph; we note that it is significantly better to start querying with the farthest nodes in the vicinity. Since the vicinities of most source-destination pairs intersect (and if they intersect, they do at least at one of the farthest nodes), queries starting from the farthest nodes achieved an improved stretch (quickly!). In terms of stretch versus query time, the results for the Internet graph were very similar to that of $G(n,m)$ graphs.

## 11 Conclusions

This paper presented data structures and query algorithms which significantly improve the space/stretch trade-off for distance oracles and compact routing schemes for the realistic case of sparse graphs. We also argued that the increased query time in our distance oracles is reasonable in practice. Allowing increased query time to improve the space/stretch trade-off brings up several interesting open problems:

- Can the query time of our schemes be reduced? In other words, can one design a distance oracle of size $O(m+n^2/\alpha)$ that returns stretch-2 paths in $o(\alpha\Delta)$ time? A more challenging problem is to design distance oracles that have size $O(m\alpha+n^2/\alpha)$ and return stretch-2 paths in $o(\alpha)$ time. We believe that the latter may require significantly new techniques.
- We presented a distributed implementation of our stretch 2 compact routing scheme only for distance oracles that have an aggregate memory requirement of $O(m\alpha+n^2/\alpha)$, but not for our linear space distance oracles (both for stretch 2 and stretch 3 schemes). While it seems significantly more challenging, a distributed version of our linear space distance oracles could have significant implications in practice: one could achieve stretch 3 with constant amount of storage at nodes in the network.
- The most intriguing problem is to compute lower bounds for distance oracles that take $\Omega(\log n)$ query time and return constant stretch paths. The holy grail of the distance oracle problem for sparse graphs is whether one can design a data structure of size $O(m \text{ polylog}(n))$ that yields constant stretch paths in $O(\text{polylog}(n))$ time. This would be a very significant result.

## References

[1] R. Agarwal, P. B. Godfrey, and S. Har-Peled. Approximate distance queries and compact routing in sparse graphs. In *Proc. IEEE International Conference on Computer Communications (INFOCOM)*, pages 1754–1762, Shanghai, China, April 2011.

[2] Y.-Y. Ahn, S. Han, H. Kwak, S. Moon, and H. Jeong. Analysis of topological characteristics of huge online social networking services. In *Proc. ACM International Conference on World Wide Web (WWW)*, pages 835–844, Alberta, Canada, May 2007.

[3] B. A. Ford. *UIA: A Global Connectivity Architecture for Mobile Personal Devices*. PhD Thesis, Massachusetts Institute of Technology, September, 2008.

[4] P. Fraigniaud and C. Gavoille. Memory requirement for universal routing schemes. In *Proc. ACM Symposium on Principles of Distributed Computing (PODC)*, pages 223–230, Ontario, Canada, August 1995.

[5] P. Fraigniaud and C. Gavoille. Local memory requirement of universal routing schemes. In *Proc. ACM Symposium on Parallel Algorithms and Architectures (SPAA)*, pages 183–188, Padua, Italy, June 1996.

[6] C. Gavoille and S. Perennes. Memory requirement for routing in distributed networks. In *Proc. ACM Symposium on Principles of Distributed Computing (PODC)*, pages 125–133, Pennsylvania, USA, May 1996.

[7] http://www.facebook.com/press/info.php?statistics.

[8] Y. Mao, F. Wang, L. Qiu, S. Lam, and J. Smith. S4: Small state and small stretch routing protocol for large wireless sensor networks. In *Proc. USENIX Symposium on Networked Systems Design and Implementation (NSDI)*, Massachusetts, USA, April 2007.

[9] M. Pătraşcu and L. Roditty. Distance oracles beyond the Thorup-Zwick bound. In *Proc. IEEE Symposium on Foundations of Computer Science (FOCS)*, pages 815–823, Nevada, USA, October 2010.

[10] M. Potamias, F. Bonchi, C. Castillo, and A. Gionis. Fast shortest path distance estimation in large networks. In *Proc. ACM Conference on Information and Knowledge Management (CIKM)*, pages 867–876, Hong Cong, China, November 2009.

[11] A. Singla, P. B. Godfrey, K. Fall, G. Iannaccone, and S. Ratnasamy. Scalable routing on flat names. In *Proc. ACM Conference on Emerging Networking Experiments and Technology (CoNEXT)*, PA, USA, December 2010.

[12] C. Sommer, E. Verbin, and W. Yu. Distance oracles for sparse graphs. In *Proc. IEEE Symposium on Foundations of Computer Science (FOCS)*, pages 703–712, Georgia, USA, October 2009.

[13] The Cooperative Association for Internet Data Analysis, 2009.

[14] M. Thorup and U. Zwick. Compact routing schemes. In *Proc. ACM Symposium on Parallel Algorithms and Architectures (SPAA)*, pages 1–10, Crete Island, Greece, July 2001.

[15] M. Thorup and U. Zwick. Approximate distance oracles. *Journal of the ACM*, 52(1):1–24, January 2005.

[16] Y. Zhang, L. Breslau, V. Paxson, and S. Shenker. On the characteristics and origins of internet flow rates. In *Proc. ACM Conference on Applications, Technologies, Architectures and Protocols for Computer Communications (SIGCOMM)*, pages 309–322, PA, USA, August 2002.